\documentclass[12pt]{article}

\usepackage{tikz}
\usetikzlibrary{matrix,arrows}
\usetikzlibrary{trees}
\usetikzlibrary{decorations.pathmorphing}
\usetikzlibrary{decorations.markings}

\usepackage{jheppub}
\usepackage{amsmath,amssymb,euscript,array,mathrsfs}
\def\Ge{G_\text{eff}}

\def\m{\mu}
\def\n{\nu}

\def\w{\omega}

\def\Im{\text{Im }}

\def\det{{\rm det}}

\def\mR{{\mathfrak R}}

\def\BB{\mathscr B}

\newcommand{\IM}{\text{Im}\,}
\def\Dbarslash{\,\,{\raise.15ex\hbox{/}\mkern-12mu {\bar D}}}
\def\Dslash{\,\,{\raise.15ex\hbox{/}\mkern-12mu D}}
\def\delslash{\,\,{\raise.15ex\hbox{/}\mkern-9mu \partial}}
\def\delbarslash{\,\,{\raise.15ex\hbox{/}\mkern-9mu {\bar\partial}}}
\def\pslash{\,\,{\raise.15ex\hbox{/}\mkern-9mu p}}
\def\aslash{\,\,{\raise.15ex\hbox{/}\mkern-9mu a}}
\def\bslash{\,\,{\raise.15ex\hbox{/}\mkern-9mu b}}
\def\cslash{\,\,{\raise.15ex\hbox{/}\mkern-9mu c}}
\def\dslash{\,\,{\raise.15ex\hbox{/}\mkern-9mu d}}

\def\AA{{\mathscr A}}

\newcommand{\MAT}[1]{\begin{pmatrix} #1\end{pmatrix}}
\newcommand{\EQ}[1]{\begin{equation}\begin{split} #1
\end{split}\end{equation}}

\newcommand{\SP}[1]{\begin{equation}\begin{split} #1
\end{split}\end{equation}}

\newcommand{\FIG}[1]{\begin{figure}\begin{center} #1 \end{center}\end{figure}}

\title{Graviton Propagation and Vacuum Polarization in Curved Space}

\author{Ross Stanley}
\author{and Timothy J. Hollowood}
\affiliation{Department of Physics, Swansea University, Swansea, SA2 8PP, U.K.}
\emailAdd{pyrs@swan.ac.uk}
\emailAdd{t.hollowood@swansea.ac.uk}

\abstract{The effects of vacuum polarization arising from loops of massive scalar particles on graviton propagation in curved space are considered. Physically, they are 
 due to curvature
induced tidal forces acting on the cloud of virtual scalar particles surrounding the graviton. The effects are tractable in a WKB and large mass limit and the results can be written as an effective refractive 
index for the graviton modes with both a real and imaginary part. The imaginary part of the refractive index is a curvature induced contribution to the wavefunction renormalization of the graviton in real affine time and can have the effect of dressing or un-dressing the graviton. The real part of the refractive index increases logarithmically at high frequency as long as the null energy condition is satisfied by the background.
}


\notoc 
\begin{document}
\maketitle

\newpage
\section{Introduction}

The propagation of quantum fields in curved spacetime is an interesting problem that is rich in surprises. For instance, it was shown by Drummond and Hathrell \cite{Drummond:1979pp} (see also \cite{Daniels:1993yi,Daniels:1995yw,Shore:1995fz,Shore:2002gn,Shore:2003zc,Shore:2007um}) that vacuum polarization is sensitive to the curvature of spacetime and leads to a modification of the wave equation equation for photons in such a way that in the low frequency limit the phase velocity can actually be greater than $c$.

The superluminal low frequency propagation of photons can be extracted from the  
effective action for QED in curved spacetime to linear order in the curvature: 
\EQ{
S&=\int d^4x\,\sqrt{-g}\Big[-\frac1{4e^2}F_{\mu\nu}F^{\mu\nu}+
a_1 R F_{\mu\nu}F^{\mu\nu}\\ &~~~~\qquad\qquad\qquad + a_2 R_{\mu\nu}
F^{\mu\lambda}F^\nu{}_\lambda+ a_3 R_{\mu\nu\lambda\rho}F^{\mu\nu}
F^{\lambda\rho}+\cdots\Big]\ .
\label{aa2}
}
The terms involving the curvature here are caused by vacuum polarization and the dots indicate that this effective action is the first term in a 
curvature/derivative expansion, so results deduced from it are valid only for
low frequency propagation. At one-loop order, the Feynman diagrams that contribute are shown in figure~\ref{f1} and it is graph (a) that is non-local and so sensitive to the background curavture.
The effective action \eqref{aa2} has been renormalized and the one-loop divergence has been absorbed into the definition of the renormalized electric charge.

The refractive index derived from \eqref{aa2}
for a photon with wave-vector $k^\mu = \w\hat k^\m$, where $\w$ is the
frequency, is
\EQ{
n_{ij}=\delta_{ij}+
\frac{\alpha}{m^2}\big(c_1\delta_{ij} R_{\mu\nu} \hat k^\mu
\hat k^\nu+c_2R_{\mu i \nu j}\hat k^\mu \hat k^\nu\big)\ ,
\label{ab}
}
where the indices $i,j$ label the two spacelike
polarisation directions. The constant coefficients $c_1,c_2$ are simply 
related to the known coefficients $a_2,a_3$ in the effective action (see for example \cite{Hollowood:2008kq}).
If we now introduce an affine coordinate $u$ along the null geodesic with tangent vector $k^\mu$, which we denote $\gamma$, we ca write the result above as
\EQ{
n_{ij}(u)=\delta_{ij}+
\frac{\alpha}{m^2}\big(c_1\delta_{ij} R_{uu}(u)
+c_2R_{iuju}(u)\big)\ .
\label{ac}
}
This makes it clear that the
Drummond-Hathrell result depends on the curvature components
$R^i{}_{uju}(u)$ and $R_{uu}(u)$ evaluated along the null geodesic $\gamma$ describing the photon's classical trajectory. 

There are three scales in the problem, $\omega$, $m$ and $\mR$, the latter being the scale of a typical element of the Riemann tensor of mass dimension $[M^{-2}]$. The Drummond-Harthrell result based on the effective action in \eqref{aa2} is valid in the limit $m^2\gg\mR$ and $\omega^2\mR/m^4\ll1$, while the
description of the photon in terms of classical ray optics in curved space requires that $\omega^2\gg\mR$, so that the wavelength is large compared with the length scale over which the curvature varies.

\FIG{
\begin{tikzpicture}
\node at (-1,3) {(a)};
\draw (2,2) circle (1cm);
\draw [fill=black] (1,2) circle (0.1cm);
\draw [fill=black] (3,2) circle (0.1cm);
\draw [decorate,decoration=snake] (-0.5,2) -- (1,2);
\draw [decorate,decoration=snake] (3,2) -- (4.5,2);
\node at (5,3) {(b)};
\draw (8,2) circle (1cm);
\draw [fill=black] (8,1) circle (0.1cm);
\draw [decorate,decoration=snake] (5.5,1) -- (10.5,1);
\end{tikzpicture}
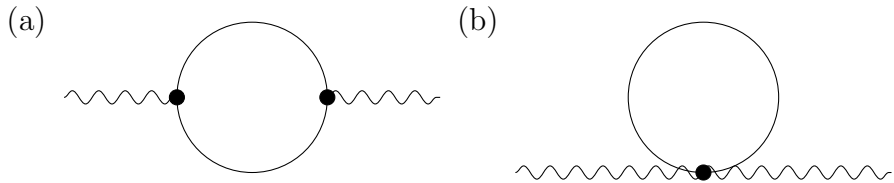
\caption{The one-loop Feynman graphs that contribute to vacuum polarization. Only the first loop (a) is non-local in spacetime and so sensitive to the background curvature.}
\label{f1} 
}

The breakthrough made in a series of papers \cite{Hollowood:2007kt,Hollowood:2007ku,Hollowood:2008kq,Hollowood:2009qz} was to extend the Drummond-Hathrell result by relaxing the requirement that $\omega^2\mR/m^4\ll1$. The result is obtained by calculating the one-loop vacuum polarization due to the Feynman graphs illustrated in Figure~\ref{f1} in the appropriate limits. The contribution from the first graph (a) is non-local in spacetime and therefore is sensitive to the background curvature. 
The effect of the non-local contribution corresponds to summing up an infinite set of terms in the effective action; however, this local expansion is generally only asymptotic and the complete result includes effects which are non-perturbative in the curvature. Note that the calculation is tractable because the two other limits, namely $\omega^2\gg\mR$ and $m^2\gg\mR$, are maintained, where the former means that we can still talk about the classical ray, or null geodesic $\gamma$, and the latter means that the result only depends on the metric in a tubular neighbourhood of $\gamma$. In fact this tubular neighbourhood can be made mathematically precise as the Penrose limit of the full metric associated to $\gamma$ \cite{Penrose} (see also \cite{Blau2,Blau:2006ar}). In terms of the so-called Brinkmann coordinates $(u,v,z^i)$ adapted to the geodesic $\gamma$, which is the curve $v=z^i=0$, the Penrose limit takes the form
\EQ{
ds^2=2du\,dv-R^i{}_{uju}(u)z^i z^j\,du^2
+dz^i\,dz^i\ .
\label{ab2}
}
The components $R^i{}_{uju}(u)$ are both the components of Riemann tensor evaluated along $\gamma$ and the non-vanishing components of the Riemann tensor of the Penrose limit itself. For later use, we also introduce the Rosen
coordinates $(u,\hat v,x^a)$, $a=1,2$, in which the Penrose limit metric
\eqref{ab2} takes the form
\EQ{
ds^2=2du\,d\hat v+C_{ab}(u)dx^a\,dx^b\ .
\label{ac2}
}
The affine parameter $u$ along $\gamma$ is common to both sets of coordinates. It turns out to be a great advantage to be able swop between the Brinkmann and Rosen coordinates when necessary. This is achieved by introducing a zweibein for the transverse part of the Rosen metric
\EQ{
C_{ab}(u)=E_{ia}(u)\delta^{ij}E_{jb}(u)\ ,
}
specifically chosen so that the tensor
\EQ{
\Omega_{ij}=\frac{dE_{ia}}{du}E_{j}{}^a
}
is symmetric. In the above $E_i{}^a$ is the inverse $E_{ia}E_j{}^a=\delta_{ij}$. In that case the relation between the two sets of coordinates is
\EQ{
\hat v=v+\frac12\Omega_{ij}(u)z^iz^j\ ,\qquad x^a=E_i{}^a(u)z^i\ .
}

The expression for the $2\times2$ matrix of refractive indices in polarisation space is \cite{Hollowood:2008kq}
\EQ{
n_{ij}(u;\omega)
=\delta_{ij}+\frac{\alpha}{2\pi\omega}\int_0^1
d\xi\,\xi(1-\xi){\cal F}_{ij}\Big(u;
\frac{m^2}{2\omega\xi(1-\xi)}\Big)\ ,
\label{ba2}
}
with
\SP{
{\cal F}_{ij}(u;z)&=\int_{-\infty+i0^+}^u\frac{du'}{(u-u')^2}\,ie^{iz(u'-u)}\,
\left[\delta_{ij}
-\Delta_{ij}\big(u,u'\big)\sqrt{\Delta(u,u')}
\right]\ ,
\label{bb2}
}
In the above $\xi$ is a Feynman parameter that is familiar in a one loop quantity.
The curvature dependence is embedded in the ``Van Vleck-Morette (VVM) matrix"
$\Delta_{ij}(u,u')$ and its determinant $\Delta(u,u')\equiv\det\,\Delta_{ij}(u,u')$. This only depends on the curvature components $R^i{}_{uju}(u)$  of the full metric evaluated along $\gamma$ and physically it encodes the tidal forces 
that are experienced along $\gamma$. In order to investigate these tidal forces, consider the geodesic equations for the transverse Brinkmann coordinates, or ``Jacobi fields",
 $z^i(u), i=1,2$,
\EQ{
\frac{d^2z^i}{du^2}+R^i{}_{uju}(u)z^j=0\ .
\label{aa}
}
The VVM matrix can then be defined in terms of the Jacobi fields as follows.  If we write the solution for the Jacobi field $z^i(u)$ in terms 
of some initial data at $u'$,\footnote{Note that the transverse Brinkmann coordinates are raised and lowered with $\delta^i{}_j$ and so one does not need to distinguish upper and lower indices.} 
\EQ{
z^i(u) =\BB_{ij}(u,u') z^j(u') + \AA_{ij}(u,u')\dot z^j(u') \ ,
\label{ad}
}
then \cite{Hollowood:2008kq}
\EQ{
\Delta_{ij}(u,u')=(u-u')\AA^{-1}_{ji}(u,u')\ .
\label{ae}
}
So the VVM matrix depends on the Jacobi fields that describe infinitesimal deformations of geodesics that pass through $u$ and $u'$ on $\gamma$. 
The VVM matrix here, is of course, specifically defined in Brinkmann coordinates. However, the determinant $\Delta(u,u')$ is a bi-scalar quantity that can be written for arbitrary coordinates as 
\EQ{
\Delta(u,u')\equiv \Delta(x(u),x(u'))\ ,\qquad\Delta(x,x')=-\frac1{\sqrt{g(x)g(x')}}
\det\,\frac{\partial^2\sigma(x,x')}{\partial x^\mu\partial x^{\prime\nu}}\ ,
\label{vvm}
}
where $\sigma(x,x')$ is the geodesic interval between points $x$ and $x'$ (defined later in \eqref{intd}). In Brinkmann coordinates it follows that
\EQ{
\Delta(u,u')=\det\,\Delta_{ij}(u,u')\ .
}
For later use, we quote an important identity \cite{Hollowood:2008kq}
\EQ{
\BB_{ij}(u,u')+\AA_{ik}(u,u')\Omega_{kj}(u')=E_{ia}(u)E_j{}^a(u')\ .
\label{nid}
}

Notice that $\Delta(u,u')$ is singular when $u$ and $u'$ are conjugate points on the geodesic $\gamma$, that is when the geodesic through $u$ and $u'$ can be infinitesimally deformed. Hence, the integral in \eqref{bb2} is only defined with an appropriate prescription which involves deforming the contour into the upper half complex plane, as indicated. The integral in \eqref{bb2} is then convergent as long as the geometry becomes flat in the asymptotic past $u'\to-\infty$. Notice that the result above for the refractive index is perfectly causal in that the refractive index at a point $u\in\gamma$ only depends on the curvature along $\gamma$ in the past $u'<u$. 

The problem before us is to consider a similar philosophy for gravitons replacing photons, interacting with some massive scalar fields. We shall assume that the number of scalar fields $N$ is large enough so that the effect of matter loops is much larger than graviton loops, although in a more complete analysis graviton loops could also be considered. 
The analogue of the terms in the QED effective action \eqref{aa2} are those involving at most two powers of the curvature:\footnote{For example, see the monograph \cite{BD}.}
\EQ{
S=\int d^4x\,\sqrt{-g}\Big[\frac1{16\pi G}(R-2\Lambda)+c_1R_{\alpha\beta\gamma\delta}R^{\alpha\beta\gamma\delta}+c_2R_{\alpha\beta}R^{\alpha\beta}
+c_3R^2+\cdots\Big]\ .
\label{lag}
}
This effective action is to be understood as being appropriately renormalized so that divergences are absorbed into the coupling $G$ and the cosmological constant $\Lambda$. In $d=4$ only two of the ${\cal O}( \mR^2)$ terms are independent as a consequence of the generalized Gauss-Bonnet theorem which implies that
\EQ{
\int d^4x\,\sqrt{-g}\big(R_{\alpha\beta\gamma\delta}R^{\alpha\beta\gamma\delta}-4R_{\alpha\beta}R^{\gamma\delta}
+R^2\big)
}
is a topological invariant whose addition to the action cannot affect the equations-of-motion. The three couplings $c_i$ depend logarithmically on the mass $m$ and the renormalization scale $\mu$. We will only need this dependence for the first two terms:
\EQ{
c_1=-\frac N{(4\pi)^2}\cdot\frac1{360}\Big[\alpha_1+\log\Big(\frac{m^2}{\mu^2}\Big)\Big]\ ,\qquad
c_2=\frac N{(4\pi)^2}\cdot\frac1{360}\Big[\alpha_2+\log\Big(\frac{m^2}{\mu^2}\Big)\Big]\ .
}
The constants $\alpha_i$ are the arbitrary finite parts of the counter terms and manifest the fact that gravity coupled to matter fields is not a perturbatively renormalizable theory and so at each order in perturbation theory new couplings appear.

In the next section, we investigate the effect of these terms of graviton propagation to find the gravitational analogue of the Drummond-Hathrell effect.

\section{The Gravitational Drummond-Hathrell Effect}

The classical propagation of gravitons in curved space is determined by expanding around a solution to Einstein's equations to linear order $g_{\mu\nu}\to g_{\mu\nu}+h_{\mu\nu}$.
As usual, one needs to fix a gauge in order to uncover the physical degree-of-freedom and to this end we impose the conventional transverse traceless gauge:
\EQ{
\nabla_\mu h^{\mu\nu}=0\ ,\qquad g_{\mu\nu}h^{\mu\nu}=0\ .
\label{gco}
}

In a general background spacetime, it is not possible to solve for the graviton modes exactly. However, just as in the QED case, we will 
work in the WKB approximation, which is valid when the
frequency is much greater than the scale over which the 
curvature varies, $\omega^2\gg\mathfrak R$.
In this case, we can write the metric perturbation in the form
\EQ{
h_{\mu\nu}(x)=\varepsilon_{\mu\nu}(x)e^{i\Theta(x)}\ ,
\label{wkb}
}
where the eikonal phase $\Theta$ is ${\cal O}(\omega)$ and the polarization tensor
$\varepsilon_{\mu\nu}$ is ${\cal O}(\omega^0)$. Substituting into the
equation-of-motion, and expanding in powers of $1/\omega$, the leading order term yields the {\it eikonal equation\/} for the phase:
\EQ{
\partial\Theta\cdot\partial\Theta=0\ ;
\label{ngr}
}
which implies that the gradient $k^\mu=\partial^\mu\Theta$ is a null vector
field. This vector field defines a {\it null congruence\/}, that is a
family of null geodesics whose tangent vectors are identified with the
vector field $k^\mu$. This vector can also be identified with the 
4-momentum of photons: in this sense the eikonal approximation is the
limit of classical ray optics.

We can then pick out a particular null geodesic, or classcal ray, in the congruence. Our goal is to find how this particular ray $\gamma$ is affected by the ${\cal O}(\mR^2)$ terms in the action. There are a set of coordinates $(u,\hat v,x^a)$ which are specifically adapted to the null congruence: these will become the Rosen coordinates of the Penrose limit
\eqref{ac2}. In these adapted coordinates the geodesics are simply lines of constant $(\hat v,x^a)$ and $u$ is the affine parameter. We will choose the ray $\gamma$ to be the one with $x^a=\hat v=0$. In these coordinates the 
eikonal phase is simply,
\EQ{
\Theta=\omega\hat v\ .
\label{ep}
}
As explained in ref.~\cite{Blau2}, the full metric
$g_{\mu\nu}$ can always be brought into the form
\EQ{
ds^2=2du\, d\hat v+C(u, \hat v,x^a)d\hat v^2+2C_a(u, \hat v,x^b)
dx^a\,d \hat v+C_{ab}(u, \hat v,x^c)dx^a\,dx^b\ .
\label{mrosen}
}

In order to find the effect of the ${\cal O}(\mathfrak  \mR^2)$ terms on the gravitation propagation we can then expand in the transverse coordinates $(\hat v,x^a)$ around $\gamma$. In fact, the terms that we need in order to fully describe the effect is precisely the Penrose limit of the metric around the null geodesic $\gamma$. This correspond to expanding in $(\hat v,x^a)$ and keeping the terms in \eqref{ac2}. The adapted coordinates are then identified with the Rosen coordinates of the plane-wave metric.
The Penrose limit has precisely the information needed to describe the behaviour of geodesics in the neighbourhood of $\gamma$ to linear order and we will see precisely encodes the information needed to investigate the effect of curvature on graviton propagation in the eikonal limit.

Once we have approximated the metric with Penrose limit, we can solve for the graviton modes exactly, these modes then correspond to the graviton modes of the full metric to leading order in the WKB approximation. There is a subtlety here in that to capture the leading order approximation involves keeping some of the components of the polarization tensor $\varepsilon_{\mu\nu}$ which are order $\omega^0$ but also other components which are order $\omega^{-1}$. This becomes clear when one solves for the exact 
complexified graviton modes in the plane-wave background:
\EQ{
h^{ab}=\frac{{\mathbb P}^{ij}E_{i}{}^a(u)E_{j}{}^{b}(u)}{\sqrt{\det\, E_{ia}(u)}}e^{i\omega \hat v}\ ,\qquad h^{\hat v\hat v}=-\frac i\omega\frac{{\mathbb P}^{ij}\Omega_{ij}(u)}{\sqrt{\det\,E_{ia}(u)}}e^{i\omega \hat v}\ ,
\label{grm}
}
with the remaining components vanishing. The two independent polarization tensors ${\mathbb P}_s$, $s=1,2$, can be chosen as 
\EQ{
{\mathbb P}^{ij}_1=\MAT{0&1\\ 1&0}\ ,\qquad {\mathbb P}^{ij}_2=\MAT{1&0\\ 0&-1}\ .
}
Notice that, as alluded to above, at leading order in the WKB approximation, we need to keep the component $h^{\hat v\hat v}$ even though it goes like $\omega^{-1}$.

In order to describe the quantum corrections to graviton propagation, one takes the graviton modes off shell by modifying the eikonal phase via a $u$ dependent factor:
\EQ{
\Theta\longrightarrow \omega\big(\hat v+\theta_{ss'}(u)\big)\ .
\label{corh}
}
The corrected phase is a $2\times2$ matrix in the two-dimensional polarization space $s=1,2$. In general, $\theta_{ss'}(u)$ will have both a real and an imaginary part. The real part describes a local correction to the phase velocity of the graviton along $\gamma$ and the imaginary part describes a local correction to the amplitude of the graviton along $\gamma$. Since $\theta_{ss'}(u)$ is perturbatively small, we can think of the result in terms of a matrix of refractive indices
\EQ{
n_{ss'}(u;\omega)=\delta_{ss'}+2\frac{d\theta_{ss'}(u)}{du}\ .
}

The one-loop quantum corrections to the equations-of-motion that are linear in the curvature come from the last three terms in \eqref{lag}.
It turns out that only the first two terms of order ${\cal O}(\mathfrak R^2)$ in \eqref{lag} contribute to graviton propagation. The Gauss-Bonnet argument can then be used to reduce this to a single term involving the square of the Ricci tensor. So for a plane wave metric the renormalized effective action to this order takes the form
\EQ{
S=\int d^4x\,\sqrt{-g}\Big[\frac1{16\pi G}\big(R-2\Lambda\big)-
\frac{N}{16\pi^2\cdot120}\lambda
R_{\mu\nu}R^{\mu\nu}\Big]\ ,
}
for a finite quantity 
$\lambda$ which is an independent dimensionless coupling constant. Substituting the mode \eqref{grm} with modified eikonal phase \eqref{corh} into the equation-of-motion that follows from the above action at leading order in the eikonal approximation yields a simple equation of the form
\EQ{
\frac{d\theta_{ss'}(u)}{du}=\frac{\lambda\Ge}{120\pi}R_{uu}(u)\delta_{ss'}\ ,
\label{gww}
}
where we have defined the effective Newton constant $\Ge=GN$.\footnote{We are assuming that $N\gg1$ so that graviton loops are suppressed relative to matter loops.}
Just as in the QED case, the result can be stated as a curvature induced contribution to the refractive index
for graviton propagation along the geodesic $\gamma$,
\EQ{
n_{ss'}(u)=\left(1+\frac{\lambda\Ge}{60\pi}R_{uu}(u)\right)\delta_{ss'}\ ,
\label{osp}
}
to ${\cal O}(\Ge)$.
The result is actually simpler than the QED case in \eqref{ac} since the result is diagonal in the 2-dimensional polarization space. 
So the effective of curvature, in the limit $\omega^2\gg{\mathfrak R}$, is to induce a frequency independent shift in the refractive index. Notice that this effect depends on the component of the Ricci tensor $R_{uu}(u)$ of the Penrose limit which is equal to the same components of the Ricci tensor of the full metric evaluated along $\gamma$.

\section{One-Loop Vacuum Polarization}
\label{SEC:candc}

The aim of this section is to compute the effect of one-loop vacuum polarization of a massive scalar field on graviton propagation in the eikonal limit $\omega^2\gg{\mathfrak R}$ and in the limit that $m^2\gg{\mathfrak R}$. This latter limit, means that the quantum fluctuations around the classical geodesic $\gamma$ are small and so allows us to approximate the metric with the Penrose limit around the geodesic $\gamma$. This leaves a non-trivial dependence on the 
dimensionless ratio $\omega^2{\mathfrak R}/m^4$.  In particular, when we talk about low and high frequency we always mean the limits:
\EQ{
\text{low:}\qquad\qquad\frac{\omega^2{\mathfrak R}}{m^4}\ll1\ ,\qquad
\text{high:}\qquad\qquad\frac{\omega^2{\mathfrak R}}{m^4}\gg1\ .
}
but subject to $\omega^2\gg\mR$.

At the one-loop level there are two graphs that contribute to the quantum equation-of-motion of the graviton field. The first is the important one from our point-of-view because it is non-local and therefore sensitive to the curvature. The second is local and we will be able to account for it by subtracting the contribution of the first in the flat space limit.
So turning to the first graph. It involves the 2-point function of the energy-momentum tensor coupled to the external gravitons. Since we are interested in the quantum corrected equation-of-motion, the relevant 2-point function is the retarded one
\EQ{
\Pi_{\m\nu\sigma\rho}(x,x')=\text{local}+\langle0|T_{\mu\nu}(x)T_{\sigma\rho}(x')|0\rangle_\text{ret}\ .
\label{laa}
}
The fact that the retarded 2-point function is needed ensures that the quantum equation-of-motion is causal, since it vanishes for $x'$ lying outside the backward lightcone of $x$. 
In the above the energy-momentum tensor of the scalar field is
\EQ{
T_{\mu\nu}=\partial_\mu\phi\partial_\nu\phi-\tfrac12g_{\mu\nu}(\partial^\rho\phi
\partial_\rho\phi+m^2\phi^2)\ .
}
The linearized quantum-corrected equation-of-motion for the graviton modes takes the form
\EQ{
\square h_{\m\n}(x)=8\pi\Ge\int d^4x'\,\sqrt{-g(x')}\,
\Pi_{\mu\nu\sigma\rho}(x,x')h^{\sigma\rho}(x')\ ,
\label{qza}
}
with, on the left-hand side, the appropriate Laplacian for gravitons in a non-trivial background.
The strategy is to solve this order-by-order in the coupling $\Ge$. At leading order, one takes the classical graviton mode \eqref{grm} with the corrected eikonal phase \eqref{corh} with $\theta$ of order $\Ge$ one the left-hand side and on the right-hand side the classical graviton mode. This leads to the equation for the one-loop ${\cal O}(\Ge)$ correction to the phase:
\EQ{
\frac{d\theta_{ss'}(u)}{du}=\text{local}+\frac{8\pi G_\text{eff}}{\omega^2}\int d^4x'\,\sqrt{-g(x')}\,h_s^{*\mu\nu}(x)
\partial_\mu\partial'_\sigma G(x,x')\,\partial_\nu\partial'_\rho G(x,x')h_{s'}^{\sigma\rho}(x')\ .
\label{mxx}
}
The insertions of the classical graviton modes involve, from \eqref{grm},
\EQ{
h_s^{\mu\nu}\partial_\mu\otimes\partial_\nu=
\mathbb P_s^{ij}\frac{e^{i\omega\hat v}}{\sqrt{\det E_{ia}}}\Big[E_i{}^aE_j{}^b\partial_a\otimes\partial_b-\frac i\omega\Omega_{ij}\partial_{\hat v}\otimes\partial_{\hat v}\Big]\ .
}
Note only the first term in the energy-momentum tensors contribute due to the gauge condition \eqref{gco}. In the above, $G(x,x')$ are the scalar propagators in the plane-wave background. These can be taken as Feynman propagators even though this does not give the retarded contribution to the 2-point function of the stress tensor required in \eqref{laa}. The reason why we can make this simplification is that we actually want the convolution of the 2-point function with the  classical graviton mode as in \eqref{qza}. The latter is a positive frequency solution with respect to the null coordinate $u$ and this has the effect of picking out the retarded part of the 2-point function of the stress tensor.\footnote{This important point is explained in detail in \cite{Hollowood:2008kq}.}

The scalar Feynman propagator in a general background spacetime can be written 
in the heat-kernel or ``proper-time'' formalism as
\EQ{
G(x,x')=\sqrt{\Delta(x,x')}
\int_0^\infty\frac{dT}{(4\pi T)^2}\,ie^{-im^2T+\tfrac i{2T}\sigma(x,x')}\Omega(x,x'|T)\ ,
\label{hyu}
}
subject to the usual $m^2\to m^2-i\epsilon$ prescription. Here, $\sigma(x,x')$ is the geodesic interval between the points $x$ and $x'$:
\EQ{
\sigma(x,x')=\frac12\int_0^1 d\tau\, g_{\mu\nu}(x)\dot x^\mu\dot x^\nu\ ,
\label{intd}
}
where $x^\mu=x^\mu(\tau)$ is the geodesic joining $x=x(0)$ and 
$x'=x(1)$. The factor $\Delta(x,x')$ 
is the famous Van Vleck-Morette determinant defined previously in \eqref{vvm}. 

The expression \eqref{hyu}
has a nice interpretation in the worldline formalism, in which 
the propagator between two points $x$ and $x'$ is determined by a sum
over worldlines $x^\mu(\tau)$ that connect $x=x(0)$ and $x'=x(T)$
weighted by $\exp iS[x]$ with the action
\EQ{
S[x]=-m^2T+\frac14\int_0^T d\tau\,g_{\mu\nu}(x)\dot x^\mu\dot
x^\nu\ .
\label{wla}
}
Here, $T$ is the worldline length of the loop which is an auxiliary
parameter that must be 
integrated over. The expression \eqref{hyu} corresponds to the
expansion of the resulting functional integral around the 
stationary phase solution, which is simply the classical 
geodesic that joins $x$ and $x'$. In particular, the classical geodesic has an action
$S[x]=\sigma(x,x')/2T-m^2T$ giving the exponential terms in
\eqref{hyu}. The VVM determinant comes from integrating
over the fluctuations around the geodesic to Gaussian order while the
term $\Omega(x,x'|T)=1+\sum_{n=1}^\infty a_n(x,x')T^n$ encodes all the
higher non-linear corrections. These terms are effectively an expansion
in $\mR/m^2$, so the form for the propagator is useful in the limit of
weak curvature compared with the Compton wavelength of the scalar particle.
Of course, this is precisely the limit we are working in here and this goes hand-in-hand with the fact that in the plane wave limit this factor is trivial $\Omega(x,x'|T)=1$.

The weak curvature 
limit $\mR\ll m^2$ leads to a considerable simplification as we now explain. Denoting the worldline lengths of each of the scalar propagators as $T_1$ and $T_2$, 
the terms in the exponent of the integrals in \eqref{mxx} are 
\EQ{
\exp\Big[-im^2T+\frac i2\Big(\frac1{T_1}+\frac1{T_2}\Big)
\sigma(x,x')+i\omega
\hat v'\Big]\ .
\label{zll}
}
For later use, we find it convenient to change variables from $T_1$
and $T_2$ to $T=T_1+T_2$ and
$\xi=T_1/T$, so $0\leq\xi\leq1$. The parameter $\xi$ is identified as a conventional Feynman parameter. Expressed the other way
\EQ{
T_1=T\xi\ ,\qquad T_2=T(1-\xi)\ .
}
The Jacobian is
\EQ{
\int_0^\infty\frac{dT_1}{T_1^2}\,\frac{dT_2}{T_2^2}=\int_0^\infty
\frac{dT}{T^3}\int_0^1\frac{d\xi}{[\xi(1-\xi)]^2}\ .
}

In the limit $\mR\ll m^2$ the integral over $x'$ is dominated by a
stationary phase determined by extremizing the exponent \eqref{zll}
with respect to $x'$:
\EQ{
\frac1{2T\xi(1-\xi)}\partial'_\mu\sigma(x,x')+
\omega\partial'_\mu \hat v'=0\ .
\label{gew}
}
Since $\partial^{\prime\mu}\sigma(x,x')$ is the tangent vector at $x'$
of the geodesic passing through $x'$ and $x$, the stationary phase solution
corresponds to a geodesic with tangent vector $\varpropto
\partial^{\prime \mu}\hat v'$. 
This means that $x$ and $x'$ must lie on a null geodesic. 
If we choose $x$ to be the point $(u,0,0,0)$ then  
$x'$ must have Rosen coordinates $(u',0,0,0)$ and so $x'$ and $x$ lie on the geodesic $\gamma$ and 
\EQ{
\partial_{\hat v'}\sigma(x,x')=u-u'\ ;
}
so the $\hat v'$ component of \eqref{gew} becomes
\EQ{
\frac{u'-u}{2T\xi(1-\xi)}+\omega=0
}
and hence
\EQ{
u'=u-2\omega T\xi(1-\xi)\ .
\label{cdf}
}

In the equivalent worldline picture, the stationary phase solution
which dominates in the limit $\mR\ll m^2$ describes a situation where
the incoming photon propgating along $\gamma$ decays to an electron positron pair at the point
$u'=u-2\omega T\xi(1-\xi)$ which then
propagate along the null geodesic $\gamma$ to the point $u$ and then
combine into the photon again. This was a key 
step in the derivation of the refractive index in the worldline formalism
in \cite{Hollowood:2007kt,Hollowood:2007ku}.

The plane-wave limit of the full metric provides precisely the data that is needed to perform the saddle-point approximation to leading order. In other words, for a plane wave background the problem is exact at Gaussian order. For the remainder of the calculation, we now simply assume a plane-wave metric knowing that the result will apply to the full metric when
$\mR\ll m^2$. The meat of the calculation involves performing the Gaussian integral over $x'$ where the exponential factor has the form \eqref{zll}. In the plane-wave background the geodesic interval has a simple form quadratic in the transverse coordinates
\SP{
\sigma(x,x')=(u-u')(\hat v-\hat v')+\tfrac12\Delta_{ab}(u,u')(x-x')^a(x-x')^b\ ,
}
where
\EQ{
\Delta_{ab}(u,u')=(u-u')\left[\int_{u'}^uC^{-1}(u'')du''\right]^{-1}_{ab}\ .
\label{ddu}
}
is the transverse part of the VVM matrix in Rosen coordinates. It is related to the same quantity in the Brinkmann coordinates by the zweibein
\EQ{
\Delta_{ij}(u,u')=E_i{}^a(u)E_j{}^b(u')\Delta_{ab}(u,u')\ .
\label{ccx}
}
The integrals over $\hat v'$ and $u'$ are trivial, the former leading to a
delta function constraint
\EQ{
\int d\hat v'\,\exp\Big[\frac{i(u'-u)\hat v'}{2T\xi(1-\xi)}
+i\omega  \hat v'\Big]=4\pi T\xi(1-\xi)\delta\big(u'-u+2\omega T\xi(1-\xi)\big)
\label{yju}
} 
which then saturates the integral over the latter. The condition on $u'$ is
precisely the saddle-point condition \eqref{cdf}.

It remains to perform  the Gaussian integrals over the two transverse coordinates 
$x^{\prime a}$.
The Gaussian integrals that we need are of the form\footnote{Where the integrals are rendered convergent by the prescription $T\to T-i0^+$ and we define $x'\cdot\Delta(u,u')\cdot x'=x^{\prime a}\Delta_{ab}(u,u')x^{\prime b}$.}
\SP{
I^{(1)}=\int d^2x'\,
e^{\tfrac i{4T\xi }x'\Delta(u,u')\cdot x'}\,e^{\tfrac i{4T(1-\xi )}x'\cdot\Delta(u,u')\cdot x'}=\frac{4iT\pi\xi (1-\xi)}
{\sqrt{\det\,\Delta_{ab}(u,u')}}\ ,
\label{int1}
}
for \eqref{mxx} with indices $\{\mu,\nu,\sigma,\rho\}=\{\hat v,\hat v,\hat v,\hat v\}$, along with
\SP{
I^{(2)}_{ab}=\int d^2x'\,\frac{\partial}{\partial x^{\prime a}}  
e^{\tfrac i{4T\xi }x'\cdot\Delta(u,u')\cdot x'}\,\frac{\partial}{\partial x^{\prime b}}e^{\tfrac i{4T(1-\xi )}x'\cdot\Delta(u,u')\cdot x'}=\frac{2\pi\xi (1-\xi)\Delta_{ab}(u,u')}
{\sqrt{\det\,\Delta_{ab}(u,u')}}\ ,
\label{int2}
}
for \eqref{mxx} with indices $\{\mu,\nu,\sigma,\rho\}=\{i,j,\hat v,\hat v\}$ or $\{\hat v,\hat v,k,l\}$, and 
\EQ{
I^{(3)}_{abcd}&=\int d^2x'\,\frac{\partial^{2}}{\partial x^{\prime a}\partial x^{\prime c}}
e^{\tfrac i{4T\xi }x'\cdot\Delta(u,u')\cdot x'}\frac{\partial^{2}}{\partial
  x^{\prime b}\partial
  x^{\prime d}}
e^{\tfrac i{4T(1-\xi )}x'\cdot\Delta(u,u')\cdot x'} \\ &=
-\frac{i\pi\xi (1-\xi)}
{T\sqrt{\det\,\Delta_{ab}(u,u')}}\big(\Delta_{ab}(u,u')\Delta_{cd}(u,u')\\ &\qquad\qquad+\Delta_{ac}(u,u')\Delta_{bd}(u,u')+\Delta_{ad}(u,u')\Delta_{bc}(u,u')\big)\ ,
\label{int3}
}
for \eqref{mxx} with indices $\{\mu,\nu,\sigma,\rho\}=\{i,j,k,l\}$.

The contribution for indices $\{i,j,k,l\}$ involves the quantity 
\EQ{
E_i{}^a(u)E_j{}^b(u)I^{(3)}_{abcd}(u,u')E_k{}^c(u')E_l{}^d(u')\ ,
}
which, using \eqref{ccx} is proportional to 
\EQ{
&E_j{}^a(u)E_{j'a}(u')\Delta_{ij'}(u,u')\Delta_{k'l}(u,u')E_{k'b}(u)E_{k}{}^b(u')\\ &\qquad\qquad+\Delta_{ik}(u,u')\Delta_{jl}(u,u')+\Delta_{il}(u,u')\Delta_{jk}(u,u')\ .
\label{nnn}
}
The contribution form the other terms involving indices $\{i,j,\hat v,\hat v\}$, $\{\hat v,\hat v,k,l\}$ and $\{\hat v,\hat v,\hat v,\hat v\}$ is proportional to
\EQ{
&\Omega_{ij}(u)I^{(2)}_{cd}(u,u')E_k{}^c(u')E_l{}^d(u')+
E_i{}^a(u)E_j{}^b(u)I^{(2)}_{ab}(u,u')\Omega_{kl}(u')\\ &\qquad\qquad
+\frac{i(u-u')^2}{4\omega T^2\xi(1-\xi)}\Omega_{ij}(u)I^{(1)}(u,u')\Omega_{kl}(u')\ ,
}
which, using \eqref{cdf}, \eqref{int1} and \eqref{int2}, is proportional to
\EQ{
\Omega_{ij}(u)\AA^{-1}_{lk'}(u',u)E_{k'b}(u)E_{k}{}^b(u')
+E_j{}^a(u)E_{j'a}(u')\AA^{-1}_{j'i}(u',u)\Omega_{kl}(u')+\Omega_{ij}(u)\Omega_{kl}(u')\ .
}
Using the identity \eqref{nid}, and being careful with the relative normalization, these terms combine nicely with the first term in \eqref{nnn}, to give a net contribution involving the tensor\footnote{Here, we use the symmetry properties $\AA_{ij}(u,u')=-\AA_{ji}(u',u)$ and $\BB_{ij}(u,u')=\BB_{ji}(u',u)$.}
\EQ{
&\big(\BB(u,u')\AA^{-1}(u,u')\big)_{ji}\big(\AA^{-1}(u,u')\BB(u,u')\big)_{lk}\\
&\qquad\qquad=\frac1{(u-u')^2}\big(\Delta(u,u')\BB^\top(u,u')\big)_{ij}(\BB^\top(u,u')\Delta(u,u')\big)_{kl}\ .
}
In particular, if we had not included the $h^{\hat v\hat v}$ term in \eqref{grm}, we would not have been able to write the final result in terms of the matrices $\AA$ and $\BB$ and, for example, the result for a symmetric plane wave to be discussed below would not have involved a function of the difference $u-u'$ which on physical grounds we expect since this particular plane wave admits a Killing vector $\partial_u$.
But with the contribution from $h^{\hat v\hat v}$, the final expression depends only on the matrices $\AA_{ij}(u,u')$ and $\BB_{ij}(u,u')$ or $\Delta_{ij}(u,u')$ that are associated to the Jacobi fields. The final tensorial nature of the result can be summarized by the quantity
\EQ{
\Delta^{(2)}_{ijkl}(u,u')&\overset{\text{def.}}=\frac1{16}\Big[(\Delta(u,u')\BB^\top(u,u'))_{ij}(\BB^\top(u,u')\Delta(u,u'))_{kl}\\ &\qquad\qquad+\Delta_{ik}(u,u')\Delta_{jl}(u,u')+\Delta_{il}(u,u')\Delta_{jk}(u,u')\Big]\ .
}
Collecting all the remaining  factors together gives the result
\EQ{
&\frac{d\theta_{ss'}(u)}{du}
=\text{local}\\ &-
\frac{\Ge}{16\pi\omega^2}\int_0^{\infty-i0^+}\frac{dT}{T^3}\int_0^1d\xi
\,e^{-im^2T}\Delta^{(2)}_{ss'}(u,u')\sqrt{\Delta(u,u')}
\Big|_{u'=u-2\omega T\xi(1-\xi)}\ ,
}
where
\EQ{
\Delta^{(2)}_{ss'}(u,u')={\mathbb P}_s^{ij}\Delta^{(2)}_{ijkl}(u,u'){\mathbb P}_{s'}^{kl}\ .
}
The local contribution comes from graph (b) in Figure~\ref{f1}. We 
can account for this contribution by subtracting the flat space limit of the non-local expression. This gives 
\EQ{
\frac{d\theta_{ss'}(u)}{du}&
=-\frac{\Ge}{16\pi\omega^2}\int_0^{\infty-i0^+}\frac{dT}{T^3}\int_0^1d\xi
\,e^{-im^2T}\\
&\qquad\qquad\times\Big[\Delta^{(2)}_{ss'}(u,u')\sqrt{\Delta(u,u')}-4\Big]_{u'=u-2\omega T\xi(1-\xi)}\ .
}
The contour for the $T$ integral is taken just below to real axis in order to avoid the singularities of the VVM determinant when $u$ and $u'$ are conjugate points.

The resulting integral is still divergent as we can see by expanding the 
integrand in powers of $\omega$. This is achieved by expanding in powers of $u-u'$:
\EQ{
\Delta^{(2)}_{ss'}(u,u')\sqrt{\Delta(u,u')}=\big(4+R_{uu}(u)(u-u')^2\big)\delta_{ss'}
+{\cal O}(u-u')^3\ .
\label{zzx}
}
This means that divergent contribution is of the form
\EQ{
\frac{d\theta_{ss'}(u)}{du}\Big|_\text{div.}=-\frac{\Ge}{120\pi}R_{uu}(u)\delta_{ss'}
\int_0^{\infty-i0^+} \frac{dT}{T}e^{-im^2T}\ .
}
We can regularize by first rotating to Euclidean proper time $T\to-iT$ and then introducing a cut off $\delta$
\EQ{
\frac{d\theta_{ss'}(u)}{du}\Big|_\text{div.}=\frac{\Ge}{120\pi}
\Big[\gamma_E+\log(m^2\delta)\Big]R_{uu}(u)\delta_{ss'}\ .
}
We can subtract the divergence and introduce an arbitrary finite piece in the form of the coupling $\lambda$. The regularized expression is then
\EQ{
\frac{d\theta_{ss'}(u)}{du}
&=\frac{\lambda\Ge}{120\pi}
R_{uu}(u)\delta_{ss'}-\frac{\Ge}{16\pi\omega^2}\int_0^\infty\frac{dT}{T^3}\int_0^1d\xi
\,e^{-im^2T}\\
&\times
\Big[\Delta^{(2)}_{ss'}(u,u')\sqrt{\Delta(u,u')}-(4+R_{uu}(u)(u-u')^2)\delta_{ss'}\Big]_{u'=u-2\omega T\xi(1-\xi)}\ ,
}
where the integrand is now regular at $u'=u$.
In the low frequency limit, we have
\EQ{
\frac{d\theta_{ss'}(u)}{du}
=\frac{\lambda\Ge}{120\pi}
R_{uu}(u)\delta_{ss'}+\frac{m^4\Ge}{\omega^2}
\,{\cal O}\left(\frac{\omega\sqrt{\mathfrak R}}{m^2}\right)^4\ ,
}
which matches the form extracted from the effective action \eqref{gww} precisely.

We can express the result above as a matrix of refractive indices and change variables from  $T$ to $u'$ to write the result in the same form as the QED result \eqref{ba2}:
\EQ{
n_{ss'}(u;\omega)
&=\delta_{ss'}-\frac{\Ge}{2\pi}\int_0^1d\xi\,[\xi(1-\xi)]^2
{\cal F}_{ss'}\Big(u;\frac{m^2}{2\omega\xi(1-\xi)}\Big)\ ,
\label{aee}
}
where
\EQ{
&{\cal F}_{ss'}(u;z)=-
\lambda R_{uu}(u)\delta_{ss'}\\ &+\int_{-\infty+i0^+}^u\frac{du'}{(u-u')^3}\,e^{iz(u'-u)}\Big[
\Delta^{(2)}_{ss'}(u,u')\sqrt{\Delta(u,u')}-(4+R_{uu}(u)(u-u')^2))\delta_{ss'}\Big]\ .
\label{twt}
}
This result is the gravitational analogue of the formula for the refractive index of the photon in \eqref{bb2}.

\section{Properties of the Gravitational Refractive Index}

In this section, we investigate some of the phenomenology of the gravitational refractive index. In order to have a concrete example to hand, we first look at the simplest kind of plane-wave spaces for which the Riemann tensor components $R^i{}_{uju}$ are constant along the geodesic. These are the symmetric plane waves, or 
Cahen-Wallach spaces \cite{Cahen}. To simplify things further, we take the case where the two eigenvalues of the Riemann tensor are equal, so up to a choice of coordinates we can write $R^i{}_{uju}=\sigma^2\delta_{ij}$, for constant $\sigma$. In this case, the VVM matrix, and related tensor $\Delta^{(2)}$, are
\EQ{
\Delta_{ij}(u,u')=\frac{\sigma(u-u')}{\sin\sigma(u-u')}\delta_{ij}\ ,\qquad
\Delta^{(2)}_{ss'}(u,u')=
\frac{4\sigma^2(u-u')^2}{\sin^2\sigma(u-u')}\delta_{ss'}\ .
\label{bw}
} 
In this case, everything is diagonal in polarization indices and so we will not show them where possible. Changing variable to $t=u-u'$, we have, from \eqref{twt},
\EQ{
{\cal F}(z)&=-
2\lambda\sigma^2+\int_0^{\infty-i0^+}
dt\,e^{-izt}\Big[\frac{4\sigma^3}{\sin^3\sigma t}-\frac4{t^3}-\frac{2\sigma^2}{t}\Big]\\
&=-2\lambda\sigma^2
-4z\sigma+\frac{11\sigma^2}3+2(z^2-\sigma^2)\Big[\log\Big(\frac{2\sigma}z\Big)+\psi\Big(\frac{3\sigma+z}{2\sigma}\Big)\Big]\ ,
\label{kaa}
}
where $\psi(x)$ is the di-gamma function. Notice that the integrand is a function of the difference $t=u-u'$ which is a special feature of a symmetric plane wave background which is translationally invariant under $u\to u+a$.

The first issue is to investigate is the behaviour of the refractive index at high and low frequencies, or more precisely large and small values of the dimensionless ration $\omega\sigma/m^2$. The low frequency behaviour is obtained by expanding \eqref{kaa} in powers of $z^{-1}$:
\EQ{
n(\omega)=1+\frac{\Ge}{2\pi}\Big(\frac{\lambda\sigma^2}{30}+\frac{17\sigma^4\omega^2}{4725m^4}+\frac{914\sigma^6\omega^4}{
945945m^8}+\cdots\Big)\ .
\label{pxx}
}
So the low frequency limit of the refractive index is independent of $\omega$ and depends on the coupling constant $\lambda$. Consequently, the 
low frequency phase velocity is not necessarily equal to $c$. As has been discussed in detail in \cite{Shore:2007um}, it is possible that $n(0)<1$ since this does not imply a violation of causality 
because low frequency waves cannot be used to send information.

Causality actually depends on the high frequency limit of the refractive index, since this is what governs the propagation of a sharp wavefront. For our plane-wave example, this can be extracted by taking the $z\to0$ limit of \eqref{kaa},
\EQ{
{\cal F}(z)\longrightarrow 2\sigma^2\log z+{\cal O}(z^0)\ .
}
Therefore, the refractive index behaves as
\EQ{
n(\omega)\ \overset{\omega\to\infty}=\  1+\frac{\Ge\,\sigma^2}{60\pi}\log\omega+\cdots\ ,
}
for large $\omega$. This indicates that at some high frequency the perturbative expansion is breaking down and --- at the very least --- one would need to sum up an infinite class of higher terms in perturbation theory in order to get a sensible result. The result above, for the high frequency behaviour is in fact universal, since the origin of the logarithm is in the divergence at the upper limit of integration of the third term in the integrand \eqref{twt} as $z\to0$. The general result for the high frequency limit, is therefore simply
\EQ{
n_{ss'}(u;\omega)\ \overset{\omega\to\infty}=\ \delta_{ss'}+\frac{\Ge}{120\pi}R_{uu}(u)\log\omega\,\delta_{ss'}+\cdots\ .
}
Interestingly, the logarithmic behaviour is increasing as long as the null energy condition is satisfied since this requires that $R_{uu}>0$.

For our simple example of a symmetric plane-wave space, the refractive index is real, however, more generally the
refractive index has both a real and an imaginary part. The real part represents a change in the phase velocity along the ray $\gamma$ while the imaginary part represents the dissipative effect of propagation caused by decay into pairs of the scalar particle, although there is nothing to prevent an amplification, as we explain below. For instance, if we expand the result \eqref{aee} in powers of the curvature by taking the next term in the series \eqref{zzx}
\EQ{
\Delta^{(2)}_{ss'}(u,u')&\sqrt{\Delta(u,u')}\\ &=\big(4+R_{uu}(u)(u-u')^2+\tfrac12\partial_uR_{uu}(u)(u-u')^3\big)\delta_{ss'}+{\cal O}(u-u')^4\ ,
\label{zzx2}
}
then one finds to order $\mR^{3/2}$,\footnote{Note that a derivative of a component of the Riemann or Ricci tensors is order $\mR^{3/2}$ in our convention.} that there is a contribution to the imaginary part of the refractive index involving the derivative of the Ricci tensor:
\EQ{
\IM\,n_{ss'}(u;\omega)=\frac{\omega\Ge}{2\pi\cdot140m^2}\cdot\partial_uR_{uu}(u)\delta_{ss'}+\cdots\ .
}
This represents a local quantum contribution to the amplitude of the graviton involving the Ricci tensor of the form
\EQ{
\exp\left[-\frac{\omega^2\Ge}{2\pi\cdot140m^2}R_{uu}(u)\right]
}
as the graviton propagates. This change in the amplitude is caused by the effect of the local curvature on the wavefunction renormalization of the graviton \cite{Hollowood:2010bd}. In order to understand this effect we have to remind ourselves that a physical graviton is a renormalized composite object that consists of a bare graviton plus a cloud of virtual scalar pairs. In flat space, there is a delicate balance between the bare graviton and the virtual cloud of scalars, but in curved space this balance is perturbed by the tidal forces acting on the composite particle induced by the curvature of spacetime.
The quantity $\Gamma=2\omega\,\IM\,n(u;\omega)$ represents the instantaneous rate of production of scalar particle pairs from the gravitons. It is 
important to realize that this can either be positive or negative as a function of $u$, since the balance between bare particle and virtual cloud can be upset either way.
In particular, in certain backgrounds the graviton can actually amplify as the virtual cloud becomes depleted. This does not break unitarity or violate the optical theorem because 
the graviton is already a renormalized particle. One way to investigate this, is to imagine turning on the coupling at finite affine time $u=u_0$. In this way, one would see the graviton undergo wavefunction renormalization even in flat space. However, in a gravitational theory, as in QED, this wavefunction renromalization is a UV sensitive process that actually diverges as the cut-off is removed.\footnote{This effect is described in detail in \cite{NEW}.} So the dressing which creates the renormalized graviton is an infinite process and the implication of this is that the graviton may become infinitely ``undressed" which means that the amplitude can increase without bound as the graviton becomes undressed by curvature induced tidal forces.
More generally the imaginary part can be written in nice way by changing variables to $t=u-u'$ and extending the contour to run from $-\infty$ to $+\infty$:
\EQ{
&\IM\,n_{ss'}(u;\omega)\\ &=\frac{i \Ge}{2\pi}\int_0^1d\xi\,[\xi(1-\xi)]^2
\int_{-\infty-i0^+}^{+\infty-i0^+}\frac{dt}{t^3}\,e^{-izt}\Delta^{(2)}_{ss'}(u,u-t)\sqrt{\Delta(u,u-t)}\ .
\label{qww}
}

A useful consistency check of the result \eqref{twt} can be obtained by taking the external graviton off-shell which is possible in a plane-wave background since it can be done  simply by modifying the eikonal phase in \eqref{ep} as follows
\EQ{
\Theta=\omega\hat v\longrightarrow\omega\hat v+\frac{M^2u}{2\omega}\ ,
}
where $M$ is the off-shell mass of the graviton. This extra phase can be carried through the one-loop computation of the graviton propagation. This extra phase only affects the non-local contribution from Feynman graph (a) in Figure~\ref{f1} and leads to a modification of the integrand in the refractive index formula in \eqref{twt} of the form\footnote{Note that the subtractions in \eqref{twt} which become the local contribution to the vacuum polarization are not modified and they are independent of $M$.}
\EQ{
\int_{-\infty+i0^+}^u\frac{du'}{(u-u')^3}\,e^{i(z-M^2/2\omega)(u'-u)}\Delta^{(2)}_{ss'}(u,u')\sqrt{\Delta(u,u')}\ .
}
The expression for the imaginary part of the refractive index in \eqref{qww} is similarly modified.
Now when $M>2m$, which corresponds to the off-shell graviton being above threshold for decay into pairs of the scalar particle, the imaginary part of the refractive index can be evaluated by completing the contour in \eqref{qww} in the upper-half $t$-plane.\footnote{More precisely only for the values of $\xi$ for which $M^2\xi(1-\xi)>m^2$.} In particular, there is a contribution from the pole at $t=0$, which is independent of the curvature, of 
\EQ{
\Im n(\omega)=\frac{m^4\Ge}{4\omega^2}\int_0^1d\xi\,\theta(M^2\xi(1-\xi)-m^2)
=\frac{m^4\Ge}{4\omega^2}\sqrt{1-4m^2/M^2}\ ,
}
diagonal in polarization indices,
which simply represents the exponential decrease in the off-shell graviton amplitude
in flat space due to decay into pairs of the scalars above threshold. The decay rate is given by $2\omega\,\Im n$ in agreement with a conventional flat space calculation of the rate.

There is another way that an imaginary part of the refractive index can be produced which is non-perturbative in the curvature. For a symmetric plane wave where
at least one of the eigenvalues of the 
constant $R^i{}_{uju}$ is negative there is a positive contribution to 
$\IM n(\w)$, so a decaying amplitude, which is
non-perturbative in the curvature. The simplest case to consider is
$R^i{}_{uju}=-\sigma^2\delta_{ij}$. Such a background violates the null energy condition, but provides a useful example where the imaginary part of the refractive index can be calculated exactly. In this case, 
$\IM n(\w)=0$ to all orders in the curvature expansion as can be seen from \eqref{pxx} by taking $\sigma\to i\sigma$. 
In this case, the VVM matrix and related tensor $\Delta^{(2)}$ are obtained by substituting $\sigma\to i\sigma$ in \eqref{bw}
\EQ{
\Delta_{ij}(u,u')=\frac{\sigma(u-u')}{\sinh\sigma(u-u')}\delta_{ij}\ ,\qquad
\Delta^{(2)}_{ss'}(u,u')=
\frac{4\sigma^2(u-u')^2}{\sinh^2\sigma(u-u')}\delta_{ss'}\ .
\label{bw2}
} 
The integral in  \eqref{qww}
can be computed by deforming the contour and picking up the contributions from the poles on the negative imaginary axis at $t=-in\pi/\sigma$, $n=1,2,\ldots$. This gives directly
\EQ{
\IM n(\w)=\frac{\Ge}{16\omega^2}\int_0^1d\xi\ \frac{m^4+4\sigma^2\omega^2(\xi(1-\xi))^2}{
1+e^{\pi m^2/2\omega\sigma\xi(1-\xi)}}\ ,
\label{by}
}
which is, indeed, non-perturbative in the curvature. In this case the
result is independent of $u$ and so it implies a constant rate of
production of pairs of the scalar particles. Such a process is kinematically
disallowed in flat space, but there is no such kinematic constraint in
curved space. Although this result has been calculated in a background that violates the null energy condition it is generic to plane waves with at least one negative eigenvalue.\footnote{We thank the referee of this paper for pointing out that Hawking radiation near the horizon of an evaporating black hole violates the null energy condition. This would be an interesting scenario in which to consider the graviton refractive index.}

To conclude it is interesting to compare the ``phenomenology" of the gravitational refractive index with the photon refractive index described in detail in 
\cite{Hollowood:2008kq,Hollowood:2009qz}. The detailed shape of the real and imaginary parts of the refractive index depends sensitively on the curvature in both cases. The obvious difference is that the photon refractive index approaches unity at high frequencies like $1/\omega$, to compare with the logarithmically diverging behaviour of the graviton case. This is reflection of the fact that QED is better behaved in the UV compared with quantum corrections to general relativity from matter fluctuations. In particular, the one-loop QED calculation is entirely self-consistent whereas the one-loop calculation in the graviton case predicts its own demise at sufficient high frequency.


\begin{thebibliography}{99}

{\small

%\cite{Drummond:1979pp}
\bibitem{Drummond:1979pp}
  I.~T.~Drummond and S.~J.~Hathrell,
  %``QED Vacuum Polarization In A Background Gravitational Field And Its Effect
  %On The Velocity Of Photons,''
  Phys.\ Rev.\ D {\bf 22} (1980) 343.
  %%CITATION = PHRVA,D22,343;%%

%\cite{Daniels:1993yi}
\bibitem{Daniels:1993yi}
  R.~D.~Daniels and G.~M.~Shore,
  %``'Faster than light' photons and charged black holes,''
  Nucl.\ Phys.\  B {\bf 425} (1994) 634
  [arXiv:hep-th/9310114].
  %%CITATION = NUPHA,B425,634;%%

%\cite{Daniels:1995yw}
\bibitem{Daniels:1995yw}
  R.~D.~Daniels and G.~M.~Shore,
  %``'Faster than light' photons and rotating black holes,''
  Phys.\ Lett.\  B {\bf 367} (1996) 75
  [arXiv:gr-qc/9508048].
  %%CITATION = PHLTA,B367,75;%%
  
%\cite{Shore:1995fz}
\bibitem{Shore:1995fz}
  G.~M.~Shore,
  %``'Faster than light' photons in gravitational fields: Causality, anomalies
  %and horizons,''
  Nucl.\ Phys.\  B {\bf 460} (1996) 379
  [arXiv:gr-qc/9504041].
  %%CITATION = NUPHA,B460,379;%%  
  
%\cite{Shore:2002gn}
\bibitem{Shore:2002gn}
  G.~M.~Shore,
  %``Faster than light photons in gravitational fields. 2. Dispersion and vacuum
  %polarization,''
  Nucl.\ Phys.\  B {\bf 633} (2002) 271
  [arXiv:gr-qc/0203034].
  %%CITATION = NUPHA,B633,271;%%

%\cite{Shore:2003zc}
\bibitem{Shore:2003zc}
  G.~M.~Shore,
  %``Quantum gravitational optics,''
  Contemp.\ Phys.\  {\bf 44} (2003) 503
  [arXiv:gr-qc/0304059].
  %%CITATION = CTPHA,44,503;%%

%\cite{Shore:2007um}
\bibitem{Shore:2007um}
  G.~M.~Shore,
  %``Superluminality and UV completion,''
  Nucl.\ Phys.\  B {\bf 778} (2007) 219
  [arXiv:hep-th/0701185].
  %%CITATION = NUPHA,B778,219;%%

%\cite{Hollowood:2007kt}
\bibitem{Hollowood:2007kt}
  T.~J.~Hollowood and G.~M.~Shore,
  %``Causality and Micro-Causality in Curved Spacetime,''
  Phys.\ Lett.\  B {\bf 655} (2007) 67
  [arXiv:0707.2302 [hep-th]].
  %%CITATION = PHLTA,B655,67;%%

%\cite{Hollowood:2007ku}
\bibitem{Hollowood:2007ku}
  T.~J.~Hollowood and G.~M.~Shore,
  %``The Refractive Index of Curved Spacetime: the Fate of Causality in QED,''
  Nucl.\ Phys.\  B {\bf 795} (2008) 138
  [arXiv:0707.2303 [hep-th]].
  %%CITATION = NUPHA,B795,138;%%

%\cite{Hollowood:2008kq}
\bibitem{Hollowood:2008kq}
  T.~J.~Hollowood and G.~M.~Shore,
  %``The Causal Structure of QED in Curved Spacetime: Analyticity and the
  %Refractive Index,''
  JHEP {\bf 0812} (2008) 091
  [arXiv:0806.1019 [hep-th]].
  %%CITATION = JHEPA,0812,091;%%

%\cite{Hollowood:2009qz}
\bibitem{Hollowood:2009qz}
  T.~J.~Hollowood, G.~M.~Shore and R.~J.~Stanley,
  %``The Refractive Index of Curved Spacetime II: QED, Penrose Limits and Black
  %Holes,''
  JHEP {\bf 0908} (2009) 089
  [arXiv:0905.0771 [hep-th]].
  %%CITATION = JHEPA,0908,089;%%

\bibitem{Penrose}
  R.~Penrose, ``{\it Any space-time has a plane wave as a limit}'', in: 
  Differential geometry and relativity, Reidel and Dordrecht
  (1976), 271-275.

\bibitem{Blau2}
  M.~Blau, ``{\it Plane waves and Penrose limits}'', Lectures given at the
  2004 Saalburg/Wolfersdorf Summer School, 
  {\tt http://www.unine.ch/phys/string/Lecturenotes.html}

%\cite{Blau:2006ar}
\bibitem{Blau:2006ar}
  M.~Blau, D.~Frank and S.~Weiss,
  %``Fermi coordinates and Penrose limits,''
  Class.\ Quant.\ Grav.\  {\bf 23} (2006) 3993
  [arXiv:hep-th/0603109].
  %%CITATION = CQGRD,23,3993;%%
  



\bibitem{NEW}
  T.~J.~Hollowood, G.~M.~Shore, ``{\it The Effect of Gravitational Tidal 
  Forces on Renormalized Quantum Fields\/}'', {to appear\/}.

\bibitem{BD}
N.~D.~Birrell and P.~C.~W.~Davies, ``{\it Quantum fields in 
curved space}'', Cambridge University Press, 1982.

\bibitem{Cahen}
M.~Cahen and N.~Wallach, ``{\it Lorentzian symmetric spaces\/}'',
Bull. Am. Math. Soc. {\bf 76} (1970) 585-591. 


%\cite{Hollowood:2010bd}
\bibitem{Hollowood:2010bd}
  T.~J.~Hollowood and G.~M.~Shore,
  %``The Effect of Gravitational Tidal Forces on Vacuum Polarization: How to
  %Undress a Photon,''
  Phys.\ Lett.\  B {\bf 691} (2010) 279
  [arXiv:1006.0145 [hep-th]].
  %%CITATION = PHLTA,B691,279;%%

  
  }

\end{thebibliography}
\end{document}